\documentclass[a4paper,11pt]{article}
\pdfoutput=1 

\usepackage{jheppub} 
                     
\usepackage{amsmath,amssymb,amsthm,amscd,graphicx}
\usepackage{psfrag}

\usepackage[english]{babel}
\usepackage[T2A]{fontenc}
\usepackage[utf8]{inputenc}
\usepackage{textcomp}
\usepackage{array}
\usepackage{hyperref}
\usepackage{tikz}
\usepackage{float}
\usetikzlibrary{decorations.markings}

\input epsf.sty

\usepackage{color}

\newcommand{\eq}[1]{\begin{equation} #1 \end{equation}}

\newcommand{\rme}{\textrm{e}}

\theoremstyle{definition}

\def\({\left(}
\def\){\right)}
\newcommand{\del}{\partial}

\newcommand{\brc}[1]{\left(#1\right)}
\newcommand{\bsq}[1]{\left[#1\right]}

\newcommand{\CN}{{\cal N}}

\newcommand{\CZ}{{\cal Z}}

\def\cO{\mathcal{O}}

\def\IN{{\mathbb N}}
\def\IZ{{\mathbb Z}}
\def\IR{{\mathbb R}}

\newcommand{\re}{{\rm e}}
\newcommand{\ri}{{\rm i}}
\newcommand{\rd}{{\rm d}}

\newcommand{\bs}{\boldsymbol}

\newcommand{\be}{\begin{equation}}
\newcommand{\ee}{\end{equation}}
\newcommand{\ba}{\begin{aligned}}
\newcommand{\ea}{\end{aligned}}
\newcommand{\ben}{\begin{eqnarray}\displaystyle}
\newcommand{\een}{\end{eqnarray}}

\newdimen\tableauside\tableauside=1.0ex
\newdimen\tableaurule\tableaurule=0.4pt
\newdimen\tableaustep
\def\phantomhrule#1{\hbox{\vbox to0pt{\hrule height\tableaurule width#1\vss}}}
\def\phantomvrule#1{\vbox{\hbox to0pt{\vrule width\tableaurule height#1\hss}}}
\def\sqr{\vbox{%
  \phantomhrule\tableaustep
  \hbox{\phantomvrule\tableaustep\kern\tableaustep\phantomvrule\tableaustep}%
  \hbox{\vbox{\phantomhrule\tableauside}\kern-\tableaurule}}}
\def\squares#1{\hbox{\count0=#1\noindent\loop\sqr
  \advance\count0 by-1 \ifnum\count0>0\repeat}}
\def\tableau#1{\vcenter{\offinterlineskip
  \tableaustep=\tableauside\advance\tableaustep by-\tableaurule
  \kern\normallineskip\hbox
    {\kern\normallineskip\vbox
      {\gettableau#1 0 }%
     \kern\normallineskip\kern\tableaurule}%
  \kern\normallineskip\kern\tableaurule}}
\def\gettableau#1{\ifnum#1=0\let\next=\null\else
\squares{#1}\let\next=\gettableau\fi\next}

\preprint{RUP-20-18}

\title{Black Hole Quasinormal Modes and Seiberg-Witten Theory}
\author[a,b]{Gleb Aminov,}
\author[c,d]{Alba Grassi}
\author[e]{and Yasuyuki Hatsuda}

\affiliation[a]{Department of Physics and Astronomy, Stony Brook University, Stony Brook, NY 11794, USA}
\affiliation[b]{ITEP NRC KI, Moscow 117218, Russia}
\affiliation[c]{Simons Center for Geometry and Physics, SUNY, Stony Brook, NY, 1194-3636, USA}
\affiliation[d]{Institut f\"{u}r Theoretische Physik, ETH Z\"{u}rich, CH-8093 Z\"{u}rich, Switzerland}
\affiliation[e]{Department of Physics, Rikkyo University, Toshima, Tokyo 171-8501, Japan}
\emailAdd{gleb.aminov@stonybrook.edu, agrassi@scgp.stonybrook.edu, yhatsuda@rikkyo.ac.jp}
\abstract{   
We present new analytic results on black hole perturbation theory.
Our results are based on a novel relation to four-dimensional $\mathcal{N}=2$ supersymmetric gauge theories. 
We propose an exact version of Bohr-Sommerfeld quantization conditions on quasinormal mode frequencies in terms of the Nekrasov partition function in a particular phase of the $\Omega$-background.
Our quantization conditions also enable us to find exact expressions of eigenvalues of spin-weighted spheroidal harmonics. 
We test the validity of our conjecture by comparing against known numerical results for Kerr black holes as well as for Schwarzschild black holes. Some extensions are also discussed.}

\begin{document}
\maketitle

\flushbottom

\section{Introduction}

Finding analytic solutions in spectral theory of quantum mechanical operators  is hard. 
Nevertheless it is recently recognized that a geometric perspective of spectral theory \cite{bpv2,voros-quartic} often provides us with many useful tools, developed in supersymmetric gauge theories \cite{ns,nrs,gmn,gmn2} and topological string theory \cite{ghm,cgm2}, to obtain exact solutions for new families of quantum spectral problems.

In this paper we apply such a geometric/gauge theoretical perspective to the spectral problems governing black hole perturbation theory.
More precisely we study so-called quasinormal modes (QNMs).
These modes do not correspond to bound states (or normal modes) but rather to resonance states (or dissipative modes) in quantum mechanics.\footnote{The mathematical origins of normal modes and of quasinormal modes are similar. Both are two-point boundary value problems.} Their spectra are discrete and complex.
The QNMs are responsible for the damped oscillations appearing, for example, in  the ringdown phase of two colliding black holes and have a direct connection to gravitational waves observations \cite{PhysRevLett.116.061102}. We refer to  \cite{review, Ferrari_2008, RevModPhys.83.793,reviewsimple,Nollert_1999} for a review on the subject and a more  exhaustive list of references.

We point out in this work that these spectral problems can be ``solved'' by using four-dimensional supersymmetric  gauge theories in a particular phase of the $\Omega$-background \cite{n, no2}. We refer to it as the Nekrasov-Shatashvili (NS) phase \cite{ns}. 
Our first step is to identify the corresponding matter contents on the gauge theory side. 
We do so by comparing a master wave equation in black hole perturbations with a differential equation originating from Seriberg-Witten (SW) theory \cite{seiberg1994, seiberg1994a}.
Interestingly, the matter contents encode the dimension of the black hole as well as  the type of asymptotic geometries.
Higher dimensional black holes are described by four-dimensional gauge theories.

In this work, we deeply look into four-dimensional asymptotically flat Schwarzschild and Kerr black holes.
We find that QNM frequencies of these black holes are determined by Bohr-Sommerfeld-type quantization conditions in $SU(2)$ SW theory with three fundamental hypermultiplets ($N_f=3$). 
Moreover, the extremal limit of the Kerr black holes turns out to correspond to the decoupling limit in SW theory, where one of the masses is sent to infinity and we are left with two fundamental hypermultiplets ($N_f=2$). 
This kind of quantization conditions has already been proposed in \cite{froman1992} for Schwarzschild black holes  in the complex WKB approach. However, the proposal in \cite{froman1992} takes the form of a formal power series in the Planck parameter. This series is known to have \textit{zero}-radius of convergence. Therefore one has to truncate the infinite sum at an optimal order as was done in \cite{froman1992} or to resum it by the Borel summation technique. We emphasize that our quantization conditions overcome this difficulty.
Our equations still have an infinite sum, but it has \textit{finite}-radius of convergence. The situation is crucially different from \cite{froman1992}.
One more advantage is that it is easy to extend it to the Kerr black holes.
We explicitly show that our quantization conditions reproduce the numerical QNM frequencies correctly.
We also present a new analytic result on the spin-weighted spheroidal eigenvalues that are eigenvalues of the angular part of the Teukolsky equation.
We find an exact expression in terms of a gauge theoretical function. 

We also have to note that there is already a similar attempt to map problems in black hole perturbation theory to those in two-dimensional conformal field theories with central charge $c=1$ \cite{novaes2014, dacunha2015, amado2017, novaes2019, barraganamado2019, dacunha2019}.  Via the Alday-Gaiotto-Tachikawa (AGT) correspondence \cite{agt}, they turn out to correspond to the graviphoton phase of the $\Omega$-background.
On the contrary our framework corresponds to the NS phase of the $\Omega$-background which, via the AGT correspondence, makes contact with two-dimensional conformal field theories where $c \to \infty$.\footnote{Surprisingly these two distinct regimes, $c=1$ and $c\to \infty$, are interrelated in an highly non-trivial way \cite{Grassi:2016nnt}, see also \cite[Sec. 6]{ggm}.}  This approach leads to a simpler and more systematic solution of the problem.  
For instance, we find a simple closed form expression of the spin-weighted spheroidal eigenvalues in terms of the Nekrasov partition function in the NS phase (see eq. \eqref{eq:A-FNS}).
In addition, we can perform explicit computations of the quasinormal modes frequencies and compare with the known numerical data. 

This paper is structured as follows.
In Sec.~\ref{general} we present the general idea behind the geometric/gauge theoretic approach to spectral theory. The main building blocks in this setup are the quantum periods which we compute explicitly by using the NS phase of the $\Omega$-background, see eqs. \eqref{Padef} and \eqref{Pdef}.

In  Sec.~\ref{flatsch} we study the simplest example: the four-dimensional asymptotically flat Schwarzschild black holes.
We propose an exact quantization condition for the corresponding QNMs frequency, see eq. \eqref{QCsch}. 
We test our proposal against available numerical data.

In  Sec.~\ref{flatkerr} we study the four-dimensional asymptotically flat Kerr black holes.
We find an explicit expression of the angular eigenvalues, see eq. \eqref{eq:A-FNS}, and we propose an exact quantization condition for the radial Teukolsky equation, see eq. \eqref{QCkerr} for the generic situation and eq. \eqref{QCkerr2} for the extremal limit.

In Sec.~\ref{other} we briefly discuss higher dimensional black holes.
We then conclude by presenting some future directions. In Appendix \ref{NSdef} we recall the definition of the NS free energy.

\section{A geometric approach to spectral theory}\label{general}
We begin with a basic review on a geometric approach to spectral theory.
Building blocks in this approach are so-called \textit{quantum periods}. Once we obtain such periods  we  can easily determine spectral quantities, 
such as the Fredholm determinant or the quantization condition.
The key idea of the geometric approach is to relate the quantum periods to gauge theoretical quantities.
The quantum periods can be introduced as follows.\footnote{See for instance \cite[Sec.~2]{ggm} for a review and a more exhaustive list of references}
Given an operator
\be \label{Ham} H(\hat x, \hat p), \quad [\hat x, \hat p]=\ri \hbar, \ee
we  associate it with a classical curve and a one form
\be \label{eq:cl} H(x,p)=E,  \qquad  \lambda(x, E)= p(x, E)\rd x ,\ee
where $x$ and $p$ are complex variables.
For a given operator \eqref{Ham}, its classical limit $\hbar \to 0$ is unique, but inversely a classical curve \eqref{eq:cl} generates an inifinite number of quantum operators. One has to fix a quantization scheme to identify a quantum operator.

In this work  we will always deal with operators \eqref{Ham} such that the corresponding classical curve \eqref{eq:cl} coincides with a SW curve of a suitable four dimensional $\CN=2$ gauge theory.  In this situation we refer to the operator \eqref{Ham} as the {\it{quantum}} SW curve.
Moreover, if the curve $ H(x,p)=E$ has genus $g$, we choose a basis of cycles 
\be \mathcal{A}_i,  \mathcal{B}_i, \quad i=1, \cdots,  g, \ee
and define the classical periods by integrating over such cycles
\be \label{clP}\Pi_{A_i,B_i}^{(0)}(E)=\oint_{\mathcal{A}_i, \mathcal{B}_i}p(x, E)\rd x, \quad i=1,\cdots, g. \ee
In the gauge theoretic framework,  \eqref{clP} are identified with  the SW periods and they encode the central charges and masses of the BPS particles in the  theory \cite{seiberg1994a,seiberg1994}. In this context $E$ is usually denoted by $u$ and it parametrises the moduli space of vacua.

Next we define WKB quantum periods  by promoting the differential $\lambda(x, E)$ to a quantum differential 
\be \lambda(x, E,\hbar)=\sum_{n\geq 0}\hbar^n Q_n(x,E)\rd x.\ee
This is formally defined in such a way that 
\be\label{wkb}  \left(H(\hat x, \hat p)-E\right)\exp\left[{\ri\over \hbar}\int^x  \lambda(y, E,\hbar) \right]=0.\ee
We then introduce WKB quantum periods by
\be\label{Pwkb} \Pi_{A_i,B_i}^{\rm WKB}(E,\hbar)=\sum_{n\geq 0}\hbar^n \oint_{\mathcal{A}_i, \mathcal{B}_i}  Q_n(x,E)\rd x = \sum_{n\geq 0}\hbar^n \pi_{A_i,B_i}^{(n)}(E). \ee
At the leading order ($n=0$), it reproduces the classical periods \eqref{clP}.
When $n$ is large one typically finds that 
\be \pi_{A_i,B_i}^{(n)}(E)\sim n !\ee
Therefore \eqref{Pwkb} has zero-radius of convergence and to define them non-perturbativelty we need to find a way to resum their $\hbar$-expansions.  The resummation problem is highly non-trivial. This is why supersymmetric gauge theory plays a crucial role in the problem: it allows to resum the $\hbar$-expansions of the WKB quantum periods into well-defined non-perturbative object: the quantum periods \cite{ns,gmn,mirmor,mirmor2}. We denote them  by
\be \Pi_{A_i,B_i}(E,\hbar). \ee
 This resummation is usually done either by using thermodynamic Bethe ansatz (TBA) equations \cite{gmn,oper} or by using instanton counting \cite{ns} (one often refers to this construction as  {\it{Bethe/gauge} }correspondence). See \cite[Sec.~3 and 4]{ggm} for a more detailed discussion. In this paper we will use the instanton counting approach.

The main examples we consider in this work are  operators which arise in the quantization of the  $SU(2)$ SW theories  with $N_f=2,3$. These were first studied in \cite{Zen11} and they read
\be \label{sc}\ba 
{\rm H}_{\rm N_f}={- \hbar^2 \frac{\rd^2}{\rd x^2}}
+{1\over 2} \overline\Lambda\brc{\rme^{x} K_{+}\Bigl( \hbar \frac{\rd}{\rd x}-u_1+ {\hbar \over 2}\Bigr)+\rme^{-x} K_{-}\Bigl(\hbar \frac{\rd}{\rd x} -u_1- {\hbar \over 2}\Bigr)}-u_0, \ea\ee
\be \overline\Lambda=
\Lambda_{N_f}^{2-N_f/2} .  \ee
The quantities $u_0$ and $u_1$ are parameters that depend on the matter content of the underlying  SW theory. We have
\be u_0=\left\{ \begin{array}{cc}
{{\Lambda}_3\over 8}  \brc{m_1+m_2+m_3}-{{\Lambda}_3^2\over 64}~, & \text{ if}\quad N_f=3~;\\
\\
{{\Lambda}_2^2\over 8}~,& \text{ if}\quad N_f=2~;
 \end{array} \right. \ee
 as well as
\be u_1=\left\{ \begin{array}{cc}
{{\Lambda}_3\over 8}~, & \text{ if}\quad N_f=3~;\\
\\
  0~, & \text{ if}\quad N_f=2~;\\
 \end{array} \right.
\ee
In SW theory 
${\bf m}=\{m_i,\cdots, m_{N_f}\}$ are the masses associated to the fundamental hypermultiplets while $\Lambda_{N_f}$ is related to the gauge coupling/dynamical scale of the theory.  
 Moreover
  \eq{K_{+}(p)=\prod_{j=1}^{N_{+}}\brc{p+ m_j}~,\quad
K_{-}(p)=\prod_{j=N_{+}+1}^{N_{f}}\brc{p+ m_j}~,}
where one can choose $N_{+}=0,..., N_f$ without loss of generality. 
In this work we chose $N_+$ such that $\rm H_{N_f}$ is a second order differential operator.  For instance, if $N_f=3$ we will take\footnote{The choice $N_+=1$ is equivalent. For $N_+=0,3$, the operator $\rm { H}_{N_f=3}$ becomes a third order differential operator. We note that even for such a third order operator, the quantum periods should finally coincide with those for $N_+=1,2$. This implies a kind of spectral dualities.} $N_+=2$.  The classical SW curve behind \eqref{sc} has genus one. 
The corresponding quantum periods are encoded by the Nekrasov-Shatashvili free energy \be  \mathcal{F}^{(N_f)}  ({a}, {\bf m},\Lambda_{N_f}, \hbar) \ee as well as the four-dimensional quantum mirror map \be  a(E, {\bf m},\Lambda_{N_f}, \hbar) .\ee 
We refer to appendix \ref{NSdef} for a definition of these quantities. 
More precisely the quantum  A period is given by 
\be  \label{Padef} \Pi_{A}^{(N_f)}(E,{\bf m}, \Lambda_{N_f},\hbar)=a(E,{\bf m},\Lambda_{N_f}, \hbar)\ee
while the quantum period B  is
\be \label{Pdef}\ba  \Pi_{B}^{(N_f)}(E,{\bf m}, \Lambda_{N_f},\hbar)=\partial_a { \mathcal{F}}^{(N_f)} ({a}, {\bf m},\Lambda_{N_f}, \hbar)\Big |_{a=a(E,{\bf m},\Lambda_{N_f}, \hbar)}.\ea \ee
The very important fact is that we have a combinatorial formula of the Nekrasov partition function \cite{n}.
It directly computes the NS free energy exactly in $\hbar$.
The quantum mirror map is also exactly related to the NS free energy.
As a consequence, the quantum periods can be exactly reconstructed by only the NS free energy.
This is a main reason why the geometric/gauge theoretical approach is so powerful in analyzing spectral theory.

According to general expectations coming from the Bethe/gauge correspondence, the discrete part of the spectrum of ${\rm H}_{\rm N_f}$  is captured by the following quantization condition 
\be \label{QC} {\Pi_{I}^{(N_f)}(E,{\bf m}, \Lambda_{N_f},\hbar)=\mathcal{N}_I \brc{n+{1\over 2}},\quad I=A, B,\quad n=0,1,2,\dots},\ee
where $\mathcal{N}_I$ is a numerical constant.
This equation has indeed a discrete set of solutions denoted by $\{E_n\}_{n\geq 0}$.
Which quantization condition, $A$ or $B$, we should impose depends on problems or on boundary conditions.
The equation \eqref{QC} is regarded as a quantum corrected Bohr-Sommerfeld rule.
In fact, at the leading order of $\hbar$, \eqref{QC} reduces to the Bohr-Sommerfeld condition because the quantum periods reduce to the classical periods.
Let us stress that  the spectral  properties of  ${\rm H}_{\rm N_f}$ clearly depend on the values of the parameters ${\bf m}, \Lambda_{N_f},\hbar$. 
On one hand one usually imposes suitable positivity conditions on    
\be\label{pos} \hbar, \Lambda_{N_f}, m_i  \ee 
so that ${\rm H}_{\rm N_f}$  has a real, discrete spectrum. In this case  one asks for eigenfunctions $\psi(x)$ of   ${\rm H}_{\rm N_f}$ to be in $ L^2(\IR)$ as in \cite{Zen11, ns}. 
On the other hand if  $ \Lambda_{N_f}, m_i, \hbar$  are complex, we  can think of the spectral problem in terms of resonances similar to what was done in \cite{gm18,Emery:2019znd,Codesido:2016aa}. 
One nice aspect of the geometric/gauge theoretic approach is that \eqref{QC} seems to be able to capture the discrete part of the spectrum independently on whether the operator is self-adjoint with normalizable eigenfunctions or not.  We refer to \cite{gm18} for a simple class of unbounded potentials studied within this framework.

Since black hole quasinormal modes are nothing but resonances, we strongly expect that their spectra are computed in the geometric framework. What we will do in the following is to reinterpret the QNM eigenvalue problem geometrically, and connect their defining equations to suitable quantum Seiberg-Witten geometries. For the  examples discussed in this paper the relevant quantum curves are \eqref{sc} with $N_f=2,3$. We will then impose the quantization condition \eqref{QC}  and check that it reproduces the correct numerical QNM frequencies as listed for instance in \cite{web}.
In order to make contact with QNMs it is useful to express \eqref{sc} in a more convenient form. 
For instance, by following    \cite{ito2017}, we can rewrite 
\be {\rm H}_{\rm N_f}\psi(x)= E \psi(x) \ee
in a form 
\begin{equation}
\begin{aligned}
\label{heunform} \left(-\hbar^2 \del_x^2+Q_{\rm N_f}(x)\right)\tilde{\psi}(x)=0.
\end{aligned}
\end{equation}
where $Q_{\rm N_f}(x)$ for $N_f=3$ is  given by
\be \ba Q_3(x)=&\frac{e^{-2 x}}{16(\sqrt{\Lambda_3} e^x-2)^2}\Big( 4 \Lambda_3+4 \Lambda_3e^{4 x} (m_1-m_2)^2+\Lambda_3e^{2 x} (\Lambda_3-24 m_3)\\
&+4 \sqrt{\Lambda_3} e^{3 x} \left(8 m_1 m_2+\Lambda_3m_3-2 \hbar ^2\right)-4 \sqrt{\Lambda_3} e^x (\Lambda_3-8 m_3)\Big)\\
&+\frac{2 E}{\sqrt{\Lambda_3} e^x-2}
\ea\ee
and
\be   \psi (x)=\exp\left[\frac{e^{-x} \left(e^x \left(\Lambda_3 x-4 (m_1+m_2+\hbar ) \log \left(2-\sqrt{\Lambda_3} e^x\right)\right)-2 \sqrt{\Lambda_3}\right)}{8 \hbar }\right]\tilde \psi (x). \ee
Let $z$ be
\begin{equation}
\begin{aligned}
z:=\frac{2 e^{-x}}{\sqrt{\Lambda_3} },
\end{aligned}
\end{equation}
and we redefine the wave function by
\begin{equation}
\begin{aligned}
\tilde \psi(x)={z}^{-1/2} \Psi(z).
\end{aligned}
\end{equation}
Then, the new function $\Psi(z)$ satisfies the wave equation in a  normal form:
\begin{equation}\label{finform}
\begin{aligned}
\hbar^2 \Psi''(z)+\widehat{Q}_3(z) \Psi(z)=0,
\end{aligned}
\end{equation}
where
\begin{equation}
\begin{aligned}
\widehat{Q}_3(z):= \frac{Q_3(x(z))}{z^2}+\frac{\hbar^2}{4z^2} =\frac{1}{z^{ 2}(z-1)^2}\sum_{i=0}^4 \widehat{A}_i z^i,
\end{aligned}
\end{equation}
with
\begin{equation}
\begin{aligned}
\widehat{A}_0&=-\frac{(m_1-m_2)^2}{4}+\frac{\hbar^2}{4},\\
\widehat{A}_1&=-E-m_1 m_2-\frac{m_3 \Lambda_3}{8}-\frac{\hbar^2}{4},\\
\widehat{A}_2&=E+\frac{3m_3 \Lambda_3}{8}-\frac{\Lambda_3^2}{64}+\frac{\hbar^2}{4},\\
\widehat{A}_3&=-\frac{m_3 \Lambda_3}{4}+\frac{\Lambda_3^2}{32},\\
\widehat{A}_4&=-\frac{\Lambda_3^2}{64}.
\end{aligned}
\label{eq:tildeA}
\end{equation}
The important observation is that the differential equation \eqref{finform} has two regular singular points at $z=0, 1$ and an irregular singular point with Poincar\'e rank one\footnote{For a second order differential equation $y''+p(z)y'+q(z)y=0$, the Poincar\'e rank $r$ at $z=\infty$ is defined by $r=1+\max(K_1, K_2/2)$ where $p(z)=\cO(z^{K_1})$ and $q(z)=\cO(z^{K_2})$ in $z \to \infty$ \cite{ronveaux1995}. If $p(z)$ is identically zero, we have $r=1+K_2/2$. 
} 
at $z=\infty$. Such a differential equation is well-known as the confluent Heun equation \cite{ronveaux1995}.
It is also well-known that the master equations in perturbations of the Kerr black holes as well as of the Schwarzschild black holes have the same singularity structure. In the proceeding sections, we will explicitly show the correspondence among the parameters. 

In the similar manner, the quantum SW curve for $N_f=2$ with $N_+=N_-=1$ leads to
\begin{equation}\label{finform2}
\begin{aligned}
\hbar^2 \Psi''(z)+\widehat{Q}_2(z) \Psi(z)=0,
\end{aligned}
\end{equation}
where
\begin{equation}
\begin{aligned}
\widehat{Q}_2(z)=-\frac{\Lambda_2^2}{16}-\frac{m_1 \Lambda_2}{2z}+\frac{4E+\hbar^2}{4z^2}-\frac{m_2\Lambda_2}{2z^3}-\frac{\Lambda_2^2}{16z^4}.
\end{aligned}
\end{equation}
This is known as the double confluent Heun equation \cite{ronveaux1995}.
We will see that the same differential equation appears from the radial part of the Teukolsky equation in the extremal limit.

%

\section{Quasinormal modes of Schwarzschild black holes} \label{flatsch}
Schwarzschild black holes are static and spherically symmetric solutions to the Einstein equation in the vacuum. The four-dimensional asymptotically flat solution is given by
\be\label{eq:sflat} {\rd } s^2=- f(r)\rd t^2 + {1\over f(r)}\rd r^2 +r^2 \left(\rd \theta^2+\sin^2\theta \rd \phi ^2\right).\ee
where
\begin{equation}
\begin{aligned}
f(r)=1-\frac{2M}{r}.
\end{aligned}
\end{equation}
Scalar ($s=0$), electromagnetic ($s= 1$) or odd-parity gravitational ($s=2$)
linear perturbations of the metric \eqref{eq:sflat} are governed by the Regge-Wheeler type equation \cite{RW}
\begin{equation}\label{eq:sch}
\begin{aligned}
\left[ f(r) \frac{d}{dr} f(r) \frac{d}{dr}+\omega^2-V(r) \right]\phi(r)=0,
\end{aligned}
\end{equation}
where $\phi(r)$ is the field encoding the radial part of the perturbation. The potential in \eqref{eq:sch} is: 
\begin{equation}\nonumber
\begin{aligned}
V(r)=f(r) \( \frac{\ell(\ell+1)}{r^2}+(1-s^2)\frac{2M}{r^3} \), \quad \ell\in \IN \quad \text{and } \ell\geq |s| .
\end{aligned}
\end{equation}
In addition the differential equation \eqref{eq:sch}  is supplied by the following quasinormal mode boundary conditions \cite{PhysRevD.1.2870}:
\be\label{BCsh} \phi(r)\sim \left\{ \begin{array}{cc}
 \re^{- \ri \omega ( r +2M\log(r-2M))} &  \text{if} \quad r\to 2M \quad \text{(ingoing)},\\ 
 \\
 \re^{ +\ri \omega ( r +2M\log(r-2M))}  & \quad \text{if} \quad r\to \infty \quad \text{(outgoing)} . 
 \end{array} \right. \ee
These boundary conditions are satisfied only for special discrete complex values of the frequency $\omega$.   

The computation of the QNM frequencies for the Schwarzschild black holes is already non-trivial.
Though there are a lot of numerical ways to compute them, analytic approaches have been less developed.
 Our goal is to compute such frequencies by using a gauge theoretical approach.
To make contact with the early result \eqref{finform}, we rewrite \eqref{eq:sch} by
\begin{equation}
\begin{aligned}
r=2Mz,\qquad \phi(r)=\sqrt{\frac{z}{z-1}} \Phi(z),
\end{aligned}
\end{equation}
and then we obtain the normal form (see for instance \cite{heun2,Fiziev:2009kh,Fiziev_2011}):
\begin{equation}
\begin{aligned}\label{RW-norm}
\Phi''(z)+ \widetilde{Q}(z) \Phi(z)=0,
\end{aligned}
\end{equation}
where
\begin{equation}\label{regf}
\begin{aligned}
\widetilde{Q}(z)&=\frac{z^2}{(z-1)^2}[(2M\omega)^2-(2M)^2 V(2Mz)]+\frac{4z-3}{4z^2(z-1)^2} \\
&=\frac{1}{z^2(z-1)^2}\sum_{i=0}^4 \widetilde{A}_i z^i,
\end{aligned}
\end{equation}
with
\begin{equation}
\begin{aligned}
\widetilde{A}_0&=-s^2+\frac{1}{4},\\
\widetilde{A}_1&=\ell(\ell+1)+s^2,\\
\widetilde{A}_2&=-\ell(\ell+1),\\
\widetilde{A}_3&=0,\\
\widetilde{A}_4&=(2M \omega)^2.
\end{aligned}
\label{eq:hatA}
\end{equation}

We now compare the equations in the form \eqref{RW-norm}-\eqref{eq:hatA} with the $SU(2)$ quantum Seiberg-Witten curve for $N_f=3$  as given in \eqref{finform}-\eqref{eq:tildeA}.
Setting $\hbar=1$, the parameter correspondence is quite simple: \begin{equation}\label{schid}
\begin{aligned}
\Lambda_3&=- 16\ri M\omega, \quad E=-\ell(\ell+1){ +}8M^2\omega^2-\frac{1}{4},\\
m_1&=s{ -}2\ri M\omega, \quad m_2=-s{ -}2\ri M\omega,\quad m_3= -2\ri M\omega.
\end{aligned}
\end{equation}
Therefore, if we think of  the Regge-Wheeler equation  from the point of view of the supersymmetric gauge theories  it is natural  to ask what is the meaning of  the quantizaton \eqref{QC} in the context of black holes. We find evidence that, by using the dictionary \eqref{schid}, the  quantization condition \eqref{QC} for the B-period indeed computes the QNM frequencies.
Our conclusion is therefore given by 
\be\boxed{\ba \label{QCsch} &\Pi_{B}^{(3)}\left(-\ell(\ell+1){ +}8M^2\omega^2-\frac{1}{4},{\bf m}, -16M\ri\omega  ,1\right)=2\,\pi\,\brc{n+{1\over 2}},\quad n=0,1,\dots,\\
&{\bf m}=\{s{ -}2\ri M\omega, -s{ -}2\ri M\omega, -2\ri M\omega\}.
\ea}\ee
For a given set of quantum numbers $\{\ell,s,n\}$, this equation admits a discrete family of complex solutions $\omega_n(\ell,s)$. 

Since the actual computation is intricate, we briefly illustrate it.
Using \eqref{Pdef}, the left hand side in the first equation of \eqref{QCsch} is expressed by the Nekrasov-Shatashvili free energy.
This free energy is computed by Nekrasov's combinatorial formula \eqref{z4d} systematically.
For $N_f=3$, we have \eqref{full} with \eqref{fi3}.
The problem is that the NS free energy includes the parameter $a$ that is not directly related to the black hole parameters.
To avoid it, we use the Matone equation \eqref{matone}. This exact relation allows us to express $a$ in terms of $E$.\footnote{This inversion is done analytically to keep track of the powers of the instanton counting parameter.}
The Matone relation is just the inverse relation of \eqref{Padef}.
Therefore we can finally eliminate $a$ from the NS free energy, and thus we can solve the quantization condition \eqref{QCsch} with respect to $M\omega$.

Recall that the WKB quantum periods \eqref{Pwkb} are formal power series in $\hbar$. Its radius of convergence is just zero.
We cannot plug $\hbar=1$ into it na\"{\i}vely.\footnote{We have to truncate the infinite sum \eqref{Pwkb} to a certain optimal order. The WKB quantization condition studied in \cite{froman1992} has this inherent problem.} On the contrary, 
 the (non-perturbatively defined) quantum periods   \eqref{Pdef} are given by the NS free energy which is \textit{exact} in $\hbar$.
All the quantum corrections are already resummed, and we can set $\hbar=1$ without any problems. 
In this sense, we refer to \eqref{QCsch} as the exact quantization condition.
However, one has to keep in mind that there still remains the sum in the instanton counting parameter $\Lambda_{N_f}$.
This sum has a finite radius of convergence, and its treatment is easier than the divergent WKB series.

We have performed the procedure above and have checked that it indeed matches the numerical values of  the Schwarzschild black hole QNMs as obtained in \cite{Berti_2006, review}\footnote{These are  nicely organised and available in \cite{web} which is our source.}.
Some examples are given in Tables  \ref{w0f}, \ref{w1f} and  \ref{w2f}. 

One  issue that we encounter in the computations is that the NS free energy \eqref{fns4d} is given by the natural series expansion in the parameter ${\Lambda_{N_f}}/a^2$. Even though this series converges, the convergence is not very fast. In that perspective it may be useful to compute the quantum periods by using TBA equations as was done in \cite{gmn, oper,ggm} instead of using the  NS free energy.\footnote{Note that the Argyres-Douglas point \cite{ard} for the $SU(2)$, $N_f=3$ SW theory is at $m_1=m_2=m_3={\Lambda}_3/8$ and $u=-E={\Lambda}_3^2/32$. This  point is quite close to \eqref{schid}  which explain at some extend  why the convergence is not very fast.}
We leave this issue as future works.

\begin{table}[h!] 
\centering
   \begin{tabular}{l l }
  \\
 Nb& $2M\omega_0$(0,0)   \\ 
\hline
 3 &$\textbf{0.2}1453301 - \textbf{0.20}342058\ri $ \\
  8 &$\textbf{0.220}88781 - \textbf{0.2097}8038 \ri $\\
12 &$\textbf{0.220909}51 - \textbf{0.209791}31 \ri$\\
 \hline
Num &    $\textbf{0.22090988}-\textbf{0.20979143} \ri $   \\
\end{tabular}    \\
\caption{ The solution $\omega_0$ to the quantization condition \eqref{QCsch} for $\ell=s=0$ and $n=0$. We denote by $\rm Nb$  the order at which we truncate the instanton counting series  ${\mathcal F}_{\rm inst}^{(3)}$ in \eqref{fi3}. We apply  Pad\'e approximants to improve the convergence of the instanton counting series.  The matching digits are shown by boldface. The numerical value is obtained from \cite{web}.  }
 \label{w0f}
 \end{table}

\begin{table}[h!] 
\centering
   \begin{tabular}{l l l}
  \\
 Nb& $2M\omega_0(1,1)$ & $2M\omega_1(1,1)$    \\
\hline
 4  & $\textbf{0.49}3115 - \textbf{0.18}0881 \ri $& $\textbf{0.4}3066732 - \textbf{0.58}87236 \ri $  \\
  8 & $\textbf{0.496}470 - \textbf{0.1849}99 \ri $& $\textbf{0.42}899228- \textbf{0.5873}530 \ri $  \\
12 & $\textbf{0.49652}6- \textbf{0.18497}4\ri  $&  $\textbf{0.429030}98 -  \textbf{0.587335}4 \ri $  \\
  \hline
Num &  $  \textbf{0.496527} -\textbf{0.184975} \ri $ & $\textbf{0.42903084} - \textbf{0.5873353} \ri $  \\
\end{tabular}    \\
\caption{ Solutions $\omega_n$ to the quantization condition \eqref{QCsch} for $\ell=s=1$ and $n=0,1$. We denote by $\rm Nb$  the order at which we truncate the instanton counting series  ${\mathcal F}_{\rm inst}^{(3)}$ in \eqref{fi3}. We apply  Pad\'e approximants to improve the convergence of the instanton counting series. The matching digits are shown by boldface. The numerical values are obtained from \cite{web}.}
 \label{w1f}
 \end{table}
\begin{table}[h!] 
\centering
   \begin{tabular}{l l l l}
  \\
 Nb& $2M\omega_0(2,2)$& $2M\omega_1(2,2)$&  $2M\omega_2(2,2)$    \\ \hline
 3 & $\textbf{0.74}80 -  \textbf{0.1}985 \ri $&   $\textbf{0.69}47713 -  \textbf{0.5}50331 \ri $& $ \textbf{0.60}0036 -  \textbf{0.95}3084 \ri $  \\
  7& $\textbf{0.74}46 -  \textbf{0.1}890 \ri $ & $\textbf{0.693}3273 -  \textbf{0.54}8018\ri $ & $  \textbf{0.6021}54 -  \textbf{0.956}237 \ri $\\
12 & $\textbf{0.747}2 -  \textbf{0.177}7 \ri $& $\textbf{0.69342}16 -  \textbf{0.5478}29 \ri  $ & $\textbf{0.60210}1 -  \textbf{0.95655}6 \ri $\\
  \hline
Num & $\textbf{0.7473} -  \textbf{0.1779} \ri $&  $\textbf{0.6934220}-\textbf{0.547830} \ri $& $\textbf{0.602107} -  \textbf{0.956554} \ri $ \\
\end{tabular}    \\
\caption{ Solutions $\omega_n$ to the quantization condition \eqref{QCsch} for $\ell=s=2$ and $n=0,1, 2$. We denote by $\rm Nb$ the order at which we truncate the instanton counting series  ${\mathcal F}_{\rm inst}^{(3)}$ in \eqref{fi3}. We apply  Pad\'e approximants to improve the convergence of the instanton counting series. The matching digits are shown by boldface. The numerical values are obtained from \cite{web}.}
 \label{w2f}
 \end{table}\newpage

\section{Quasinormal modes of Kerr black holes} \label{flatkerr}

Kerr black holes are stationary and axially symmetric solutions to the Einstein equation in the vacuum. The  four-dimensional asymptotically flat solution in the Boyer-Lindquist coordinates is:
\be \ba \rd s^2=&-\rd t^2 +\rd r^2 +2 \alpha \sin^2 \theta \rd r \rd \phi+(r^2+\alpha^2 \cos^2\theta ) \rd \theta^2+(r^2 +\alpha^2)\sin^2\theta \rd\phi ^2  \\
&+{2 M r\over r^2+\alpha^2 \cos^2\theta}\left( \rd t +\rd r+\alpha\sin^2\theta \rd \phi \right)^2~,
\ea\ee
where $M$  is the mass and $\alpha$ is the angular momentum. 
Perturbations of rotating black holes  are described by the Teukolsky equation \cite{Teu,Teu1}.
The Tuekolsky equation is a separable partial differential equation in the Boyer-Lindquist coordinates.

After separation of variables, its {\it angular} part reads (see for instance \cite[eq.~(25)]{review})
\begin{equation}
\begin{aligned}
\biggl[ \frac{d}{dx}(1-x^2)\frac{d}{dx}+(cx)^2-2csx+{}_sA_{\ell m}+s-\frac{(m+sx)^2}{1-x^2} \biggr] {}_sS_{lm}(x)=0,
\end{aligned}
\label{eq:angular}
\end{equation}
where $x=\cos \theta$ and $s$ is the (minus of) spin of a perturbing field. Moreover  
\be \ell = 0, 1, 2\cdots,   \quad \text{with}\quad  |m|\leq \ell ,\ee 
where $ m \in \IZ$ for integer spins and  $ m \in {1\over 2}+\IZ$ for half integer spins. 
 In the black hole perturbation, the parameter $c$ is related to the angular momentum $\alpha$
and the frequency $\omega$ by \be c=\alpha\omega. \nonumber\ee
The eigenfunction ${}_sS_{lm}(x)$ is called the spin-weighted spheroidal harmonics in the literature.
Its eigenvalue ${}_sA_{\ell m}$ is determined by the regularity condition of ${}_sS_{lm}(x)$ at $x=\pm 1$.
For general $s$, $l$, $m$ and $c$, no closed form of ${}_sA_{\ell m}$ is known so far. However, for $c=0$ the  spheroidal harmonics  ${}_sS_{lm}(x)$  reduces to the   spin-weighted spherical harmonics ${}_sY_{lm}$ and
 one has
\be{}_sA_{\ell m}(c=0)=\ell(\ell+1)-s(s+1). \ee

The {\it radial} Teukolsky equation is more complicated and reads (see for instance \cite[eq.~(25)]{review}),
\begin{equation}
\begin{aligned}
\Delta(r) R''(r)+(s+1)\Delta'(r) R'(r)+V_T(r) R(r)=0,
\end{aligned}
\label{eq:radial}
\end{equation}
where $\Delta(r)=r^2-2Mr+\alpha^2$. The potential is
\begin{equation}
\begin{aligned}
V_T(r)=\frac{K(r)^2-2is(r-M)K(r)}{\Delta(r)}-{}_sA_{\ell m}+4is\omega r+2\alpha m\omega-\alpha^2\omega^2,
\end{aligned}
\end{equation}
where $K(r)=(r^2+\alpha^2)\omega-\alpha m$.
Note that the radial differential equation \eqref{eq:radial} has (regular) singular points at $r=r_\pm:=M\pm \sqrt{M^2-\alpha^2}$ corresponding to the Cauchy and event horizons. In addition, \eqref{eq:radial} is supplied by the following boundary conditions (see for instance \cite[eq. 80]{review})
\be R(r)\sim \left\{ \begin{array}{cc}
  (r_+-r_-)^{-1-s+\ri \omega+\ri \sigma_+} \re^{\ri \omega r_+}(r-r_+)^{-s-i\sigma_+} & \quad \text{if} \quad r\to r_+ ~,\\ 
 \\
 A(\omega)r^{-1-2s+\ri \omega }\re^{\ri \omega r}\quad\quad\quad\quad\quad \quad\quad\quad\quad\quad& \quad \text{if} \quad r\to \infty ~,
 \end{array} \right. \label{eq:BCKerr} \ee
 where  \be \sigma_+={\omega r_+-{\alpha m\over 2M}\over\sqrt{1-{\alpha^2\over M^2}}}.\ee
Both the angular and the radial parts of the Teukolsky equation have the same singularity structure as the confluent Heun equation, see for instance   \cite{heun2,Fiziev:2009kh,Fiziev_2011}.

For the angular part, we change the variable $z=(1+x)/2$, and define $y(z):=\sqrt{1-x^2}{}_sS_{lm}(x)/2$.
Then we obtain
\begin{equation}
\begin{aligned}
y''(z)+Q(z) y(z)=0,
\end{aligned}
\label{eq:radial-2}
\end{equation}
where $Q(z)$ takes the form
\begin{equation}
\begin{aligned}
Q(z)=\frac{1}{z^2(z-1)^2} \sum_{i=0}^4 A_i z^i~.
\end{aligned}
\label{eq:Q}
\end{equation}
The coefficients in $Q(z)$ are computed straightforwardly.
Similarly, defining $z=(r-r_-)/(r_+-r_-)$ and $y(z):=\Delta(r)^{(s+1)/2} R(r)$ for the radial part,
we obtain the same form as \eqref{eq:radial-2} and \eqref{eq:Q} with different coefficients.
Hence, the Teukolsky equation also has a connection with the quantum Seiberg-Witten geometry with gauge group $SU(2)$ and $N_f=3$ hypermultiplets. To find the precise dictionary we need to compare \eqref{eq:radial-2} and \eqref{eq:Q} with \eqref{finform}-\eqref{eq:tildeA}.

For the {\it angular} part, we find 
\begin{equation} 
\begin{aligned}
\Lambda_3&=16c,\quad E=-{}_sA_{\ell m}-s(s+1)-c^2-\frac{1}{4}, \\
m_1&=-m, \quad m_2=m_3=-s.
\end{aligned}
\label{eq:id0}
\end{equation}

For the {\it radial} part, we have
\begin{equation} 
\begin{aligned}
\Lambda_3&=-16\ri \omega \sqrt{M^2-\alpha^2}, \\
E&=-{}_sA_{\ell m}-s(s+1)+(8M^2-\alpha^2)\omega^2-\frac{1}{4}, \\
m_1&=s { -}2\ri M\omega,\qquad m_3=-s{ -}2\ri M\omega, \\
m_2&=\frac{\ri(-2M^2\omega-\alpha m)}{\sqrt{M^2-\alpha^2}}. 
\end{aligned}
\label{eq:id}
\end{equation}
When $\alpha=0$, it reproduces the identification in the Schwarzschild case by exchanging $m_2 \leftrightarrow m_3$.
This relabelling comes from the fact that the Teukolsky equation at $\alpha=0$ does not take the form of the Regge-Wheeler equation.

Notice that the extremal limit in Kerr black holes corresponds to $\alpha\to M$.
 Given the above dictionary,  this translates into
 \be {\Lambda}_3\rightarrow 0,\quad  {m_2\rightarrow\infty} , \quad  {{\Lambda}_3\, m_2=-16M\omega {(2M\omega-m)}} \quad \text{ fixed }. \ee
 From the gauge theory point of view this is precisely the  decoupling limit under which the $N_f=3$ theory flows to the $N_f=2$. 
 Hence the radial part of the Teukolsky equation in the  extremal limit corresponds to the quantum Seiberg-Witten curve of SQCD with $N_f=2$, provided we use the following dictionary
 \begin{equation} 
\begin{aligned}
\Lambda_2^2&=-16M\omega {(2M\omega-m)}, \\
E&=-{}_sA_{\ell m}^{\rm}-s(s+1)+7M^2\omega^2-\frac{1}{4}, \\
m_1&=s{-}2\ri M\omega,\qquad m_2=-s{-}2\ri M\omega. 
\end{aligned}
\label{eq:idext}
\end{equation}
This result is also obtained by starting with \eqref{eq:radial} in the extremal case $\alpha=M$.
By changing the variable $r=\sqrt{M(2M \omega-m)/\omega}\, z+M$, the radial equation is finally written as the normal form of the double confluent Heun equation which we can compare with \eqref{finform2}.

The identifications  \eqref{eq:id0}  and \eqref{eq:id} allow us to find equations determining both ${}_sA_{\ell m}$ and $\omega$ by using quantities in $\mathcal{N}=2$ SW theory. We will demostrate this in the next two subsections.

\subsection{Exact quantization condition for the angular Teukolsky equation}
Note that in the angular equation \eqref{eq:angular} or the identification \eqref{eq:id0}, the multipole number $\ell$ does not appear explicitly.
In the perspective of the quantization conditions, this must appear as a quantum number.
It turns out that the angular eigenvalues are exactly determined by the A-period quantization condition:
\begin{equation}\label{A-QC}
\begin{aligned}
\Pi_A^{(3)}\(-{}_sA_{\ell m}-s(s+1)-c^2-\frac{1}{4}, {\bf m}, 16c, 1\)=\ri \(\ell+\frac{1}{2}\),
\end{aligned}
\end{equation}
where ${\bf m}=\{-m,-s,-s\}$.
The difference between the A-period condition here and the B-period condition in the previous section (and also in the next subsection) is explained as follows. In the angular problem we impose the boundary conditions at the two regular singular points $z=0,1$.
In the radial problem, we have to impose the conditions at the regular singular point $z=1$ and at the irregular singular point $z=\infty$.
Therefore we have to consider different period integrals in the WKB approximation.

Now we rewrite the conjecture \eqref{A-QC} in a more elegant form. 
The condtion \eqref{A-QC} is nothing but the quantization condition for $a$ (see \eqref{Padef}).
Also, recall the Matone relation \eqref{matone}, which is the inverse  of \eqref{Padef}.
Therefore, we finally conclude that the equation \eqref{A-QC} is equivalent to
\begin{equation}
\label{eq:A-FNS}
\boxed{\ba
&{}_sA_{\ell m}-{}_sA_{\ell m}^{(0)}+c^2
=\Lambda_3 \frac{\del \mathcal{F}_{\rm inst}^{(3)}}{\del \Lambda_3}(\ri \ell+{\ri /2}, {\bf m},\Lambda_3,1 )\Big|_{ \Lambda_3=16c}\\
& {\bf m}=\{-m,-s,-s\},\\ 
\ea}\ee
where ${}_sA_{\ell m}^{(0)}=\ell(\ell+1)-s(s+1)$ and $\mathcal{F}_{\rm inst}^{(3)}$ is defined in Appendix \ref{NSdef}.
The first few terms read
\be\ba
{}_sA_{\ell m}-{}_sA_{\ell m}^{(0)}+c^2&=-\frac{2 c m s^2}{L^2}+{2 c^2\over L^6 (4 L^2-3)}\Big(L^6 \left(L^2+m^2-1\right)\\
&\quad+s^4 \left((5L^2+3) m^2-3L^4\right)+2L^4 s^2 \left(L^2-3 m^2\right)\Big)+\mathcal{O}(c^3),
\ea
\end{equation}
where $L^2=\ell(\ell+1)$, and we have used $ \mathcal{F}_{\rm inst}^{(3)}$ defined in \eqref{fi3}. 
The identity \eqref{eq:A-FNS} can be compared with the small-$c$ expansion of ${}_sA_{\ell m}$ in \cite{seidel1989, berti2006a} up to $c^6$.
Note that the similar consideration is found in \cite{dacunha2019}, but \eqref{eq:A-FNS} looks simpler and more direct. The expression of ${}_sA_{\ell m}$ up to $\mathcal{O}(c^{12})$ can be found in the attached \textit{Mathematica} file.

\subsection{Exact quantization condition for the radial Teukolsky equation}
The story for the radial equation is the same as for the Schwarzschild case.
The radial equation has a discrete set of complex frequencies $\omega_n(\ell,s,m)$, which have been computed numerically for instance in  \cite{Berti_2006, review, Richartz:2015saa}. 
From a gauge theoretic perspective, and  given the identification \eqref{eq:id}, it is natural to conjecture that  the frequencies $\omega_n(\ell,s,m)$ can be obtained
by imposing the following B-period quantization condition
\be \label{QCkerr} \boxed{\ba& \Pi_{B}^{(3)}\left(-{}_sA_{\ell m}-s(s+1)+(8M^2-\alpha^2)\omega^2-\frac{1}{4}  ,{\bf m},-16\ri \omega \sqrt{M^2-\alpha^2},1\right) =2\,\pi\brc{n+{1\over 2}},\quad n\geq 0\\
&{\bf m}=\biggl\{s{-}2\ri M\omega,\frac{\ri(-2M^2\omega-\alpha m)}{\sqrt{M^2-\alpha ^2}},-s{-}2\ri M\omega \biggr\} ,\ea}\ee
where ${}_sA_{\ell m}$ is as in \eqref{eq:A-FNS} while $\Pi_{B}^{(3)}$ is defined in \eqref{Pdef}. We checked that the solutions  $\omega_n(\ell,s,m)$ to \eqref{QCkerr} indeed reproduce the correct QNM's frequencies as computed numerically in \cite{Berti_2006, review} (we took the data from \cite{web}).  An example is given in Table \ref{egkerr}.

In the extremal limit, we can get a simplified quantization which now involves the quantum period of the $N_f=2$ theory and reads  
\be \label{QCkerr2}\boxed{ \ba& \Pi_{B}^{(2)}\left(-{}_sA_{\ell m}^{\rm}-s(s+1)+7M^2\omega^2-\frac{1}{4},{\bf m}, -{4 \ri \sqrt{\omega {(2M^2\omega-M m)}}} ,1\right)=2\,\pi\brc{n+{1\over 2}},\quad n\geq 0\\
&{\bf m}=\{s{-}2\ri M\omega,-s{-}2\ri M\omega \} ,\ea}\ee
where   $\Pi_{B}^{(2)}$ is defined by \eqref{Pdef}.
We check that the quantization condition \eqref{QCkerr2}  reproduces the correct numerical QNM's frequencies in the extremal case.  Some examples are given in Table \ref{egextr}.
\begin{table}[h!] 
\centering
   \begin{tabular}{l l l}
  \\
 Nb& $M\omega_0$ &  $M\omega_1$\\
\hline
 3 &\textbf{0.1}073438 ~ - \textbf{0.10}16159 \ri  & \textbf{0.08}9515 - \textbf{0.3}30273 \ri  \\
 8 &\textbf{0.1105}221~ - \textbf{0.104}7959 \ri  & \textbf{0.086}036 - \textbf{0.347}811 \ri  \\
11 &\textbf{0.110533}0 - \textbf{0.104801}3\ri & \textbf{0.0862}16 - \textbf{0.3476}86 \ri  \\
 \hline
Num &  \textbf{0.1105331} -  \textbf{0.1048015} \ri &  \textbf{0.086203} - \textbf{0.347664} \ri 
 \\
\end{tabular}    \\
\caption{Solutions $\omega_n$ to the quantization condition  \eqref{QCkerr} of Kerr BH for  ${ \alpha\over M}={1\over 10}$ with $\ell=s=m=0$ and quantum number $n=0,1$. We denote by $\rm Nb$ the order at which we truncate the   instanton counting series   ${\mathcal F}_{\rm inst}^{(3)}$ in \eqref{fi3}. The matching digits are shown in boldface. The numerical values are obtained from \cite{web}.  }
 \label{egkerr}
 \end{table}

\begin{table}[h!] 
\centering
   \begin{tabular}{l l l}
  \\
 Nb& $M\omega_0$  &  $M\omega_1$\\
\hline
2 &\textbf{0.1}0626294 - \textbf{0.08}8291652 $\ri $ &0.071401 - \textbf{0.31}74727 \ri  \\
 6 &\textbf{0.11024}328 - \textbf{0.0894}30682 $\ri $ &\textbf{0.062}368 - \textbf{0.318}6182 \ri\\
10&\textbf{0.110245}45 - \textbf{0.089433}151 $\ri $ & \textbf{0.0623}53 - \textbf{0.31884}23 \ri  \\ 
12&\textbf{0.110245}48 - \textbf{0.089433}184 $\ri $ & \textbf{0.0624}72 - \textbf{0.31884}36 \ri  \\
 \hline
Num & \textbf{0.110245}\quad~~- \textbf{0.089433} \ri & \textbf{0.062473} - \textbf{0.318840} \ri \\
\end{tabular}    \\
\caption{Solutions $\omega_n$ to the quantization condition \eqref{QCkerr2} of Kerr BH in the extremal limit  with $M=\alpha$  for $\ell=s=m=0$ and quantum number $n=0,1$. We denote by $\rm Nb$ the order at which we truncate the instanton counting series   ${\mathcal F}_{\rm inst}^{(2)}$ in \eqref{fi2}. We also use Pade approximant to accelerate the convergence. The matching digits are shown in boldface. The numerical values are obtained from  \cite{Richartz:2015saa}.  Note that in the extremal limit the numerical values are not as precise as in the general case and it seems that for $\omega_0$ we already get a few additional digits as compared to \cite{Richartz:2015saa}. }
 \label{egextr}
 \end{table} \newpage

\section{Simple extensions}\label{other}
In the above sections we  focused  on four-dimensional black holes with asymptotic flatness. 
Our gauge theoretic approach is not restricted to these particular examples.
Here we briefly illustrate some other examples.
More detailed analysis will be reported elsewhere.

A lesson we have learned from the previous examples is that it is important to understand the singularity structure of differential equations.
The singularity information tells us the matter contents of gauge theories.
Our strategy is the following. We first read off the singularity structure of master wave equations for various black holes. 
Next we look for gauge theory counterparts by comparing the Riemann sphere with punctures associated with quiver gauge theories. 
Such punctures describe singularities of quantum SW curves.
They were analyzed for regular punctures by Gaiotto \cite{gaiotto2012}, and generalized for irregular punctures by several people \cite{gaiotto2013, cecotti2011, Bonelli:2011aa}.
After identifying a gauge theory, the third task is to find relations between the parameters. This last step is quite complicated.

\subsection{Four-dimensional Schwarzschild (A)dS black holes}
A simple extension is four-dimensional asymptotically (A)dS black holes.
It is known that the master equations of these black holes have four regular singular points \cite{suzuki1998}.
This differential equation is well-known as the Heun equation. Here
we  quickly look at it for the Schwarzschild (A)dS$_4$ case.
The higher-dimensional case is discussed in the next subsection.
The radial master equations in the scalar/electromagnetic/odd-parity gravitational perturbations take the same form as \eqref{eq:sch}, but the functions are now modified as 
\be f(r)=1-\frac{2M}{r}-\frac{\Lambda}{3} r^2~, \ee
and
\be V(r)=f(r)\biggl[ \frac{\ell(\ell+1)}{r^2}+(1-s^2)\biggl( \frac{2M}{r^3}-\frac{4-s^2}{6} \Lambda \biggr) \biggr], \ee
where $\Lambda$ is the cosmological constant.
After the redefinition of $\phi(r)=\Phi(r)/\sqrt{f(r)}$, we get the normal form:
\begin{equation}
\begin{aligned}
\Phi''(r)+q(r) \Phi(r)=0,
\end{aligned}
\label{eq:normal-dS}
\end{equation}
where
\begin{equation}
\begin{aligned}
q(r)=\frac{\omega^2-V(r)}{f(r)^2}+\frac{f'(r)^2}{4f(r)^2}-\frac{f''(r)}{2f(r)}.
\end{aligned}
\end{equation}
The algebraic equation $f(r)=0$ has three roots in general. 
It is easy to see that these three points as well as $r=0$ are regular singular points of the master equation.
The infinity point is subtle.
To see the behaviour near $r=\infty$, it is convenient to change the variable $y=1/r$ and $\Phi(r)=\widetilde{\Phi}(y)/y$, and we find
\begin{equation}\label{eq:norm-ads}
\begin{aligned}
\widetilde{\Phi}''(y)+\frac{q(1/y)}{y^4} \widetilde{\Phi}(y)=0.
\end{aligned}
\end{equation}
We look at the behaviour of $q$  in the limit $y \to 0$:
\begin{equation}
\begin{aligned}
\frac{q(1/y)}{y^4}=-\frac{(1-s^2)(4-s^2)}{2y^2}+(\text{regular terms}), \quad y \to 0.
\end{aligned}
\end{equation}
It is obvious that for $s=1,2$, the infinity $r=\infty$ is not a singular point.
For $s=0$ however it is a regular singular point.\footnote{Here we are considering minimally coupled massless scalars perturbations. In  \cite{suzuki1998} the Authors study instead conformally coupled scalar perturbations and  they found that the singular point at the infinity is removable. We also remark that the regular singular point $z=\infty$ in the minimally coupled massless scalar perturbations is actually a so-called apparent (or false) singularity.}
We conclude that for the scalar perturbation the master radial equation has five regular singular points,
while for the electromagnetic and the odd-parity gravitational perturbations they have four regular singular points.
In both cases, all the singular points are regular, and thus the differential equations are Fuchsian.

For $s=1,2$ the master equation is equivalent to the Heun equation. It turns out that this case corresponds to the gauge theory with four fundamental matters ($N_f=4$).
The detailed analysis in this case will be reported soon \cite{agh2}.
The singularity structure for $s=0$ is realized in an $SU(2)\times SU(2)$ quiver gauge theory \cite{gaiotto2012}.
We do not explain it any more in this work.

\subsection{Higher-dimensional extensions}
Let us proceed to higher-dimensional cases.
For simplicity, we focus on asymptotically (A)dS Schwarzschild black holes in $d$-dimension.
The metric in this geometry is
\begin{equation}\label{line}
\begin{aligned}
\rd s^2=-f(r)\rd t^2+\frac{\rd r^2}{f(r)}+r^2 \rd\Omega_{d-2}^2,
\end{aligned}
\end{equation}
where $\rd\Omega_{d-2}^2$ is the line element of the $(d-2)$-dimensional unit sphere $S^{d-2}$.
The function $f(r)$ takes the form 
\begin{equation}
\begin{aligned}
f(r)=1-\(\frac{r_0}{r} \)^{d-3} -\frac{2\Lambda}{(d-2)(d-1)} r^2,
\end{aligned}
\end{equation}
where $r_0^{d-3}$ is proportional to the mass with a non-trivial coefficient.
We consider the higher-dimensional analog with the odd-parity gravitational perturbation of Regge and Wheeler.
In the terminology of Kodama and Ishibashi \cite{kodama2003}, it corresponds to the vector-type gravitational perturbation.
The master equation in this case is again the same as \eqref{eq:sch} with the potential \cite{kodama2003}:
\begin{equation}
\begin{aligned}
V(r)=f(r) \biggl[ \frac{(2\ell+d-4)(2\ell+d-2)}{4r^2}-\frac{3(d-2)^2 r_0^{d-3}}{4r^{d-2}}-\frac{d-4}{2(d-1)}\Lambda \biggr].
\end{aligned}
\end{equation}
For $d=4$, it actually reduces to the one for $s=2$ in the previous subsection.

We look at singularities of the master equation.
If $\Lambda \ne 0$, the equation $f(r)=0$ has $d-1$ roots.
These are regular singular points in general.
It is also easy to check that $r=0$ is a regular singular point.
To see the behavior at $r=\infty$, we rewrite the master equation in the form \eqref{eq:norm-ads}.
The coefficient function has the Laurent expansion:
\begin{equation}
\begin{aligned}
\frac{q(1/y)}{y^4}=-\frac{(d-4)(d-2)}{4y^2}+(\text{regular terms}), \quad y \to 0.
\end{aligned}
\end{equation}
We conclude that the infinity $r=\infty$ is a regular singular point except for $d=4$. We observe that for $d>4$, $r=\infty$ is actually an apparent singularity.  For $d>4$ and $\Lambda \ne 0$, the master equations are thus Fuchsian differential equations with $d+1$ singular points.
The quantum SW curve having the same singular structure is an $SU(2)^{d-2}$ quiver gauge theory \cite{gaiotto2012}.
It is interesting to see that the dimensional information of the black hole is reflected in the number of quiver gauge groups.

For the flat case $\Lambda=0$, the equation $f(r)=0$ has $d-3$ roots. These as well as $r=0$ are regular singular points.
It turns out that the infinity $r=\infty$ is an irregular singular point with Poincar\'e rank one.
Therefore in this case, we have to consider $SU(2)^{d-3}$ quiver theories associated with the Riemann sphere with an irregular puncture \cite{gaiotto2013, cecotti2011, Bonelli:2011aa}.

\section{Outlook}

Inspired by  recent developments in the gauge theoretical approach to spectral theory, we analysed the black hole QNMs in this framework. This approach provided us with some new analytic results on the QNM frequencies. We pointed out that their master equations can be written as quantum SW curves and we obtained an exact, analytic expression for their quantization condition.  We mostly focused on four-dimensional asymptotically flat Schwarzschild and Kerr black holes and we check our results against available numerical data.  We also presented a preliminary analysis 
for  asymptotically (A)dS case, which is connected to $N_f=4$, as well as the higher dimensional examples.
In this situation  it would be interesting to investigate how the existence of unstable BH solutions is reflected on the SW theory side\footnote{We would like to thank Martin Ro\v cek   for a discussion on this point.}. A more detailed study will appear elsewhere. 
Likewise, even though  in this work we focused on uncharged black holes, we expect our analysis to carry on for Reissner-Nordstr{\"o}m and Kerr-Newman black holes as well\footnote{At least for the cases where the master equation is separable.}.

In our approach a key role is played by the quantum periods of the underlying SW theory which we computed by using the Nekrasov--Shatashvili free energy  \eqref{fns4d}.  Even though  this quantity is a convergent series in the  instanton counting parameter, the convergence is a bit slow. From that perspective it would be good to compute these quantum periods by using the alternative TBA approach as  in  \cite{gmn, oper,ggm}. 
Likewise it would be interesting to  study the singularities in the Borel plane and see if there is any interpretation as  wall  crossing phenomena from the black hole viewpoint.  

Although we have been focusing on the quantization condition for the QNMs frequencies, the geometric/gauge theoretic approach to spectral theory also allows for the computation of the eigenfunctions. It would be important to pursue this direction in more details and eventually provide a more rigorous derivation of the quantization condition proposed in this paper.

Our result also indicates that it should be possible to compute efficiently the WKB expansion for the QNMs by using the holomorphic anomaly equation for the underlying gauge theory, similar to what was done in \cite{cm-ha,coms, huang}. It would be interesting to investigate this aspect more in detail as a possible alternative to the existing approaches, see for instance  \cite{Iyer:1986np, Matyjasek_2017,  BLOME1984231, Ferrari:1984aa, Hatsuda_2020}.

An additional interesting point is the connection with the work of \cite{Motl:2003cd, Natario:2004jd, Berti:2003aa,Berti:2003ab,Berti_2004,Keshet_2008} on the highly damped  QNMs. From our perspective the large $n$ behaviour of the frequencies $\omega_n$ is encoded in the asymptotics of the quantum periods/Nekrasov--Shatashvili partition function which is accessible analytically.  We hope to report on this in the near future.

\acknowledgments{
We would like to thank Masashi Kimura, Marcos Mari\~no, Martin Ro\v cek  and Ricardo Schiappa for valuable discussions and correspondence.
The work of G.A is supported by the "BASIS" Foundation, Grant No. 18-1-1-50-3.
The work of A.G. is partially supported by SNSF, Grant No. 185723.
The work of Y.H. is supported by JSPS KAKENHI Grant No. JP18K03657.
}

\appendix
\section{The Nekrasov-Shatashvili free energy}\label{NSdef}
Below we review the $U(2)$ Nekrasov-Shatashvili free energy  with $N_f$ flavours \cite{n,no2,ns,Flume:2002az}, we mostly follow the notation of \cite{ggm}.  Let us denote by
\be Y=(y_1, y_2, \cdots),\ee 
 a Young Tableau (or partition) and by
\be Y^t=(y_1^t, y_2^t, \cdots),\ee
 its transposed. We  use
\begin{equation}
  \label{Y-vec}
  \boldsymbol{Y}=(Y_1, Y_2)
\end{equation}
to denote a vector of Young tableaux and define
\begin{equation}\label{ly}
  \ell( \boldsymbol{Y})= \sum_{I=1}^2 \ell(Y_I) \ ,
\end{equation}
where
\begin{equation}
  \ell(Y)= \sum_i y_i \ .
\end{equation}
Given a  Young tableaux $Y$ and a  box $ s=(i,j) $  we define
\begin{equation}
  h_Y(s)=y_i-j, \qquad v_Y(s)= y^t_j-i.
\end{equation}
The four dimensional   $U(2)$ Nekrasov partition function with $N_f$ fundamental hypermultiplets is then defined as
\be
\label{z4d}
Z^{(N_f)} ({a};{\bf m}; \Lambda_{N_f},\epsilon_1, \epsilon_2)=\sum_{\boldsymbol{Y}} \Biggl({\Lambda_{ N_f}^{(4-N_f)}\over 4}\Biggr)^{\ell (\boldsymbol{Y})}  \CZ_{\boldsymbol{Y}}^{\rm gauge} \CZ_{\boldsymbol{Y}}^{\rm matter}, 
\ee
where  ${\bf m}=\{m_1, \cdots, m_{N_f}\}$ and
\be
\ba
\label{zdefgauge}\CZ_{\bs{Y}}^{\rm gauge} &=\prod_{I,J=1}^2 \prod_{s \in Y_I} {1\over \alpha_I -\alpha_J -\epsilon_1 v_{Y_J}(s) +\epsilon_2 \left( h_{Y_I}(s)+1 \right)} \\
&\qquad \qquad \times  \prod_{s\in Y_J}
{1\over \alpha_I -\alpha_J +\epsilon_1 \left(v_{Y_I}(s)+1\right){-} \epsilon_2 h_{Y_J}(s)},
\ea
\ee
with $    \alpha_2=-  \alpha_1=a $ .
Likewise
\be
\ba
\label{zdefmatter}\CZ_{\bs{Y}}^{\rm matter} &=\prod_{k=1}^{N_f}\prod_{I=1}^2 \prod_{(i,j) \in Y_I} \left(\alpha_I +m_k+\left(i-\frac{1}{2}\right)\epsilon_1+\left(j-\frac{1}{2}\right)\epsilon_2\right)\\
\ea
\ee
The instanton part of the Nekrasov-Shatashvili (NS) free energy is defined by \cite{ns} \begin{equation}
\label{fns4d}
F_{\rm inst}^{(N_f)}({a}; {\bf m} ;  \Lambda_{N_f},\hbar) = -\hbar \, \lim_{\epsilon_2
  \rightarrow 0} \, \epsilon_2 \log Z^{(N_f)}({ \ri}{a},   {\bf m} ,  \hbar , \epsilon_2).
\end{equation}
An important property of Nekrasov partition function \eqref{fns4d}, \eqref{z4d} is that it is exact in $\hbar$, $\epsilon_i$  and it is a {\it{convergent}} series in $\Lambda_{N_f}/a^2$, see for instance \cite{ilt,felder,bsu}.

For the theory with $N_f=3$ we have ${\bf m}=\{m_1,m_2,m_3\}$
and  the first few terms read
\be\label{mf3} \ba F_{\rm inst}^{(3)}({a};  {\bf m};\Lambda_{3} , \hbar) = &\frac{1}{8} \left(-\frac{4 m_1 m_2 m_3}{4 a^2+\hbar ^2}+m_1+m_2+m_3+{  \hbar} \right)\Lambda_3 \\
&+{ \Lambda_3 ^2 \over 4096}\left( \frac{192 m_1^2 m_2^2 m_3^2}{a^2 \left(4 a^2+\hbar ^2\right)^2}+\frac{1024 m_1^2 m_2^2 m_3^2}{\left(4 a^2+\hbar ^2\right)^3}-\frac{\left(a^2+4 m_1^2\right) \left(a^2+4 m_2^2\right) \left(a^2+4 m_3^2\right)}{a^4 \left(a^2+\hbar ^2\right)}\right.\\
&\left. +\frac{16 \left(a^2 \left(m_1^2 \left(m_2^2+m_3^2\right)+m_2^2 m_3^2\right)+4 m_1^2 m_2^2 m_3^2\right)}{a^4 \left(4 a^2+\hbar ^2\right)}+5\right)
+\mathcal{O}(\Lambda_3^3) 
\ea \ee
Likewise when $N_f=2$ we have  ${\bf m}=\{m_1,m_2\}$ and
\be \ba F_{\rm inst}^{(2)}({a};  {\bf m}; \Lambda_{2}, \hbar) =& \left({1\over 2}- \frac{ \left(4 m_1 m_2\right)}{2 \left(4 a^2+\hbar ^2\right)}\right){\Lambda_2^2\over 4} -\\
&{\Lambda_2^4\over 1024 \left(a^2+\hbar^2\right) \left(4 a^2+\hbar^2\right)^3}\Big(64 a^2 \left(a^4+3 a^2 \left(m_1^2+m_2^2\right)+5 m_1^2 m_2^2\right)+\hbar ^6\\
&12 \hbar ^4 \left(a^2+m_1^2+m_2^2\right)+16 \hbar ^2 \left(3 a^4+6 a^2 \left(m_1^2+m_2^2\right)-7 m_1^2 m_2^2\right) \Big)+\mathcal{O}\left(\Lambda_2 ^6\right)\ea\ee
To obtain the partition function for the $SU(2)$ theory one has to divide the $U(2)$ partition function by the $U(1)$ factor \cite{agt}. In our conventions this translates into  a small modification in the 1st instanton factor.
For $N_f=3$ we have 
\be \label{fi3} {\mathcal{F}}_{\rm inst}^{(3)}({a};  {\bf m};  \Lambda_{3}, \hbar) = F_{\rm inst}^{(3)}({a};  {\bf m};\Lambda_{3} , \hbar) -\frac{\Lambda_3}{8} \left(+m_1+m_2+m_3+{  \hbar} \right) \ee
Likewise when $N_f=2$ we have
\be \label{fi2}{ {\mathcal{F}}}^{(2)}_{\rm inst}({a}; {\bf m};\Lambda_{3}, \hbar) =  F^{(2)}_{\rm inst}({a};  {\bf m};\Lambda_{2} , \hbar)-{\Lambda_2^2\over 8}\ee
We  define the full NS free energy as 
\be \label{full} \ba \partial_a { \mathcal{F}}^{({N_f})} ({a}; {\bf m};\Lambda_{N_f}, \hbar) =& -2\,a\brc{4-N_f}\log\bsq{\frac{\Lambda_{N_f} 2^{-{1\over (2-N_f/2)}}}{\hbar }}-\pi\hbar-2\,\ri\,\hbar\log\bsq{
\dfrac{\Gamma\brc{1+\frac{2\ri a}{\hbar}}} {\Gamma\brc{1-\frac{2\ri a}{\hbar}}}}\\
&-\ri\,\hbar
\sum_{j=1}^{N_f}\log\bsq{\dfrac{\Gamma\brc{\frac12+\frac{m_j-\ri a}{\hbar}}}
{\Gamma\brc{\frac12+\frac{m_j+\ri a}{\hbar}}}} +\dfrac{\partial {\mathcal F}_{\rm inst}^{(N_f)}({a};    {\bf m}  ; \Lambda_{N_f}, \hbar) }{\partial a}. \ea\ee 
In the context of the Bethe/gauge correspondence one also uses Matone relation \cite{matone,francisco,lmn,bkk-matone}
\be\label{matone}{E=a^2-\dfrac{\Lambda_{N_f}}{4-N_f} \dfrac{\partial \mathcal{F}_{inst}^{(N_f)}(a;{\bf m};\Lambda_{N_f}, \hbar)}{\partial \Lambda_{N_f}}.}\ee 
This relation can be inverted leading to the so-called four dimensional quantum mirror map
\be \label{mirror}a({E}; {\bf m};\Lambda_{N_f}, \hbar).\ee
This terminology comes from the fact that the identity \eqref{mirror} is a particular limit of the quantum mirror map appearing in toric Calabi-Yau manifolds \cite{acdkv}. The quantity $a$  is essentially the Kahler parameter while $E$ is the complex modulus. 

\bibliographystyle{JHEP}
\bibliography{biblio}
\end{document}